\documentclass[amsmath, amsfonts, superscriptaddress, showpacs, twocolumn, prb]{revtex4}
\usepackage{graphicx}
\usepackage{epsfig}
\pdfoutput=1
\usepackage{bm}
\usepackage{dcolumn}
\usepackage{amsmath}
\usepackage{amssymb}
\usepackage{euscript}
\renewcommand{\Re}{\mathrm{Re}}
\renewcommand{\Im}{\mathrm{Im}}

\begin{document}

\title{Effect of SDW
fluctuations on the specific heat jump in iron pnictides at a
superconducting transition}
\author{D.~Kuzmanovski}
\affiliation{Department of Physics, University of Wisconsin,
Madison, Wisconsin 53706, USA}

\author{A.~Levchenko}
\affiliation{Department of Physics and Astronomy, Michigan State
University, East Lansing, Michigan 48824, USA}

\author{M.~Khodas}
\affiliation{Department of Physics and Astronomy, University of
Iowa, Iowa City, Iowa 52242, USA}

\author{M.~G.~Vavilov}
\affiliation{Department of Physics, University of Wisconsin,
Madison, Wisconsin 53706, USA}

\begin{abstract}
Measurements of the specific heat jump at the onset of
superconducting transition in the iron-pnictide compounds revealed
strong variation of its magnitude as a function of doping that is
peaked near the optimal doping. We show that this behavior is direct
manifestation of the coexistence between spin-density-wave and
superconducting orders and the peak originates from thermal
fluctuations of the spin-density-waves near the end point of the
coexistence phase -- a tetracritical point.  Thermal fluctuations
result in a power-law dependence of the specific heat jump that is
stronger than the contribution of mass renormalization due to
quantum fluctuations of spin-density-waves in the vicinity of the
putative critical point beneath the superconducting dome.
\end{abstract}

\date{\today}
\pacs{74.70.Xa, 74.25.Bt, 74.40.-n, 74.40.Kb}
\maketitle

\section{Introduction}
The concept of quantum criticality is at the forefront of the
physics of strongly correlated materials.~\cite{Sachdev} The
discovered superconductivity in the iron-pnictide
compounds~\cite{FeSC-1,FeSC-2,FeSC-3,FeSC-4} that emerges in the
close proximity to magnetic
instability~\cite{PD-Exp-1,PD-Exp-2,PD-Exp-3,PD-Exp-4} provides new
opportunities to study quantum critical phenomena in the system with
multiple order parameters. The observed microscopic coexistence
between spin-density-wave (SDW) and superconducting (SC) orders in
iron-pnictide superconductors
(FeSC)~\cite{SC-SDW-Coex-1,SC-SDW-Coex-2,SC-SDW-Coex-3,SC-SDW-Coex-4,
SC-SDW-Coex-5,SC-SDW-Coex-6,SC-SDW-Coex-7,SC-SDW-Coex-8,SC-SDW-Coex-9,
SC-SDW-Coex-10,SC-SDW-Coex-11} implies that SDW transition line
extends into the superconducting state. If this line reaches zero
temperature the system develops quantum critical point (QCP) beneath
superconducting dome. Such a scenario is further complicated by the
fact that besides the SDW transition, there is also a nematic
transition, below which the tetragonal symmetry of the system is
spontaneously broken down to an
orthorhombic.~\cite{Nematic-1,Nematic-2,Nematic-3} The transition
line of the nematic order also enters the superconducting dome which
may lead to yet another QCP. A magnetic QCP without
superconductivity and in the case of nodal fermions in $d$-wave
superconductors has been a subject of intensive study and is known
to give rise to non-Fermi liquid behavior, and to singularities in
various thermodynamic and transport
characteristics.~\cite{QCP-1,QCP-2,QCP-3,QCP-4} The multi-band
unconventional superconductivity in FeSC brings new intriguing
questions concerning the role of QCP in thermodynamic and transport
properties of correlated
materials.~\cite{QCP-Lambda,QCP-FeSC-Review,QCP-C,QCP-Nematic,AL-Lambda,
Debanjan-Lambda,Takuya-Lambda,QCP-FeSC-Fernandes}

Recently we have received compelling experimental evidence that
superconductivity in FeSCs indeed hosts quantum criticality. Low
temperature measurements of the doping dependence of the London
penetration depth $\lambda(x)$ in clean samples of isovalent
BaFe$_2$(As$_{1-x}$P$_x$)$_2$ revealed sharp peak in $\lambda(x)$
near the optimal doping $x_c\simeq0.3$.~\cite{QCP-Lambda}
Magneto-oscillations data show an increase in effective mass
$m^*(x)$ on one of the electron Fermi surfaces as $x$ approaches
$x_c$.~\cite{QCP-dHvA} Nuclear magnetic resonance (NMR) experiments
show that the magnetic ordering temperature approaches zero at
$x_c$.~\cite{QCP-NMR} Specific heat jump $\Delta C$ displays
nonmonotonic dependence on $x$ when measured across $x_c$ at the
superconducting critical temperature $T_c$.~\cite{QCP-C} Finally,
the system obeys linear temperature dependence of the resistivity
close to $x_c$.~\cite{QCP-Transport} Observation of the same set of
features in the other families of FeSCs has been elusive so far
since quantum critical effects are easily masked by inhomogeneity
and impurity scattering. BaFe$_2$(As$_{1-x}$P$_x$)$_2$ is
particularly useful in this regard since the substitution of As by
the isovalent ion P does not change the electron/hole balance and
does not induce appreciable scattering unlike in the electron-doped
Ba(Fe$_{1-x}$Co$_x$)$_2$As$_2$ compound.~\citep{AL-Lambda}

Measurements of the magnitude and the doping dependence of the
specific heat jump were instrumental for determining and
understanding the phase diagram of iron--pnictide superconductors.
Experiments~\cite{C-jump-1, C-jump-2, C-jump-3} revealed that
$\Delta C/T_c$ vary greatly between underdoped
Ba(Fe$_{1-x}$Ni$_x$)As$_2$ and optimally hole-doped
Ba$_{1-x}$K$_x$Fe$_2$As$_2$, but even for the given material, e.g.
Ba(Fe$_{1-x}$Co$_x$)As$_2$ or BaFe$_2$(As$_{1-x}$P$_x$)$_2$, the
value of $\Delta C/T_c$ has its maximum near the optimal doping and
then decreases, approximately as $\Delta C/T_c\propto T^2_c$ at
smaller and larger dopings. It is useful to recall that in BCS
theory specific heat jump $\Delta C/T_c=4\pi^2N_F/7\zeta(3)$ is
universally determined by the total quasiparticle density of states
$N_F$ at the Fermi surface. The origin of the strong doping
dependence of $\Delta C(x)$ was rooted~\cite{MV-C} to the
coexistence of SDW magnetism and $s^{\pm}$ superconductivity and the
mean field theory is in general consistent with experimental
observations. However, the sharply peaked and highly nonmonotonic
variation of $\Delta C/T_c$ near $x_c$ as seen in the
experiment~\cite{QCP-C} is beyond the mean field treatment and is
clearly related to fluctuation effects.

Combined accurate data analysis~\cite{QCP-Lambda,QCP-C,QCP-dHvA} on
the magneto-oscillations, specific heat jump and magnetic
penetration depth near the optimal doping lead to the conjecture
that the quasiparticle mass renormalization expected close to a QCP
is the main factor which is causing the observed sharp features.
Although this is certainly the case for the explanation of the
low-temperature $\lambda(x)$ measurements, we take the point of view
that interpretation of the $\Delta C(x)$ data obtained near the
critical temperature requires an account of thermal fluctuations.

In this work, we find that thermal SDW fluctuations lead to a
dominant contribution to the specific heat jump at the onset of
superconducting transition that scales as a power law $\Delta
C/T_{c} \propto \vert x - x_c \vert^{-\alpha}$. The value of the
exponent $\alpha = 1 \div 3/2$ depends on whether SDW transition is
commensurate or incommensurate. We recall that in the 122-family of
iron-pnictides, and possibly in other FeSCs, optimal doping $x_c$
nearly coincides with the end point of the coexistence region -- a
tetracritical point $P$. Once the system is tuned to the proximity
of the tetracritical point both SDW and SC order parameters develop
strong fluctuations. In the quantum case of $T = 0$, when the whole
FS, except possibly for isolated hot points, is gapped by the
non-zero SC order parameter $\Delta \neq 0$, fluctuation effects are
reduced.\cite{AL-Lambda}  On the contrary,  near $T = T_c$, the SC
order parameter vanishes $\Delta = 0$, and SDW fluctuations are not
suppressed, giving rise to large thermal corrections.

The rest of the paper is organized as follows. In Sec.~II we
introduce the minimal two-band model of FeSCs and discuss emergent
phase diagram at the mean field level. In Sec.~III we incorporate
fluctuation effects and compute renormalized free energy of the
system. In Sec.~IV we use the latter to address the scaling of the
specific heat jump near the tetracritical point and compare our
calculations to the recent experimental findings. In Sec.~V we
summarize our main results and draw final conclusions.

\section{Model of F$\mathrm{e}$SC and phase diagram}
We consider the minimal two-band low-energy model consisting of one
circular hole pocket near the center of the Brillouin zone (BZ) and
an electron pocket near its corner.~\cite{MV-model1,MV-model} Away
from the perfect nesting electron-like band can be parametrized as
follows $\xi_e=-\xi_h+ 2 \delta_{\phi q}$, where hole band
dispersion is assumed quadratic $\xi_h=\mu_h-p^2/2m_h$, with
$\delta_{\phi
q}=\delta_0+\delta_2\cos(2\phi)+(v_Fq/2)\cos(\phi-\phi_0)$. The
parameter $\delta_{\phi q}$ captures the relative shift in the Fermi
energies, and difference in effective masses of the electron and
hole bands, via $\delta_0$, and an overall ellipticity of the
electron band, via $\delta_2$.~\cite{MV-model1} In addition,
$\delta_{\phi q}$ also captures the incommensurability of the SDW
order with vector  $\bm{q}$, where $\phi$ and $\phi_0$ are the
directions of Fermi velocity $\bm{v}_F$ and $\bm{q}$ respectively.
For isovalent doping ($\textrm{As} \rightarrow \textrm{P}$) both
$\delta_0$ and $\delta_2$ change, as the shape of the bands changes
with doping $x$. Earlier calculations show~\cite{MV-model} that
there is a broad parameter range $\delta_2/\delta_0$ for which SDW
order emerges gradually, and its appearance does not destroy SC
order; i.e., SDW and SC orders coexist over some range of dopings.
For simplicity, in our analysis we assume that only $\delta_0$
changes, while the ellipticity parameter $\delta_2$ is fixed,
although the picture is expected to stay similar for different
choices of dependence of $(\delta_{0}, \delta_{2})$ on doping. The
incommensurability vector $\bm{q}$ is an adjustable parameter that
minimizes the system free energy in the SDW phase or describes
inhomogeneous SDW fluctuations in non-magnetic  phases.

The basic Hamiltonian for electron-electron interaction includes the
free fermion part, and four-fermion interaction terms. The
interaction terms in the band basis are Hubbard, Hund, and
pair-hopping interactions, dressed by coherence factors from the
diagonalization of the quadratic form. There are five different
interaction terms in the band basis:~\cite{AC-model} two
density-density intra-pocket interactions (these interactions are
often treated as equal), density-density inter-pocket interaction,
exchange inter-pocket interaction, and inter-pocket pair hopping.
These five interactions can be rearranged into interactions in the
particle-particle channel, and spin- and charge-density-wave
particle-hole channels. For repulsive interactions, SDW and SC
channels are the two most relevant ones. We decompose these
four-fermion interactions by using SDW and SC order parameters
$\bm{M}_{q}$ and $\Delta$, and express corresponding couplings in
terms of the bare transition temperatures $T_{c0}$ to the SC state
in the absence of SDW and $T_{s0}$ to the perfectly nested FS in the
absence of SC. Thus we arrive at the following free energy density:
\begin{eqnarray}\label{F-GL}
\hskip-.25cm \frac{\mathcal{F}(\Delta,\bm
M_q)}{N_F}=\frac{\Delta^2}{2}\ln\left(\frac{T}{T_{c0}}\right)+\frac{|\bm{
M}_q|^2}{2}\ln\left(\frac{T}{T_{s0}}\right)\nonumber\\
\hskip-.25cm -2\pi
T\!\!\sum_{\varepsilon_n>0}\!\!\left[\Re\left\langle\sqrt{\mathcal{E}^2_n+|\bm{
M}_q|^2}\right\rangle_\phi\!\!\!-\varepsilon_n-\frac{\Delta^2+|\bm{
M}_q|^2}{2\varepsilon_n}\right],
\end{eqnarray}
where $\langle\ldots\rangle_\phi$ denotes averaging over $\phi$
along Fermi surfaces, $\mathcal{E}_n=E_n+i\delta_{\phi q}$,
$E_n=\sqrt{\varepsilon^2_n+\Delta^2}$, and $\varepsilon_n=\pi
T(2n+1)$ are the fermionic Matsubara frequencies
($n=0,\pm1,\pm2,\ldots$). In Eq.~\eqref{F-GL} we allowed $\bm{M}_q$
to be a vector that has freedom in orientation as well as in the
choice of the nesting vector $q$.

\begin{figure*}
  \includegraphics[width=15cm]{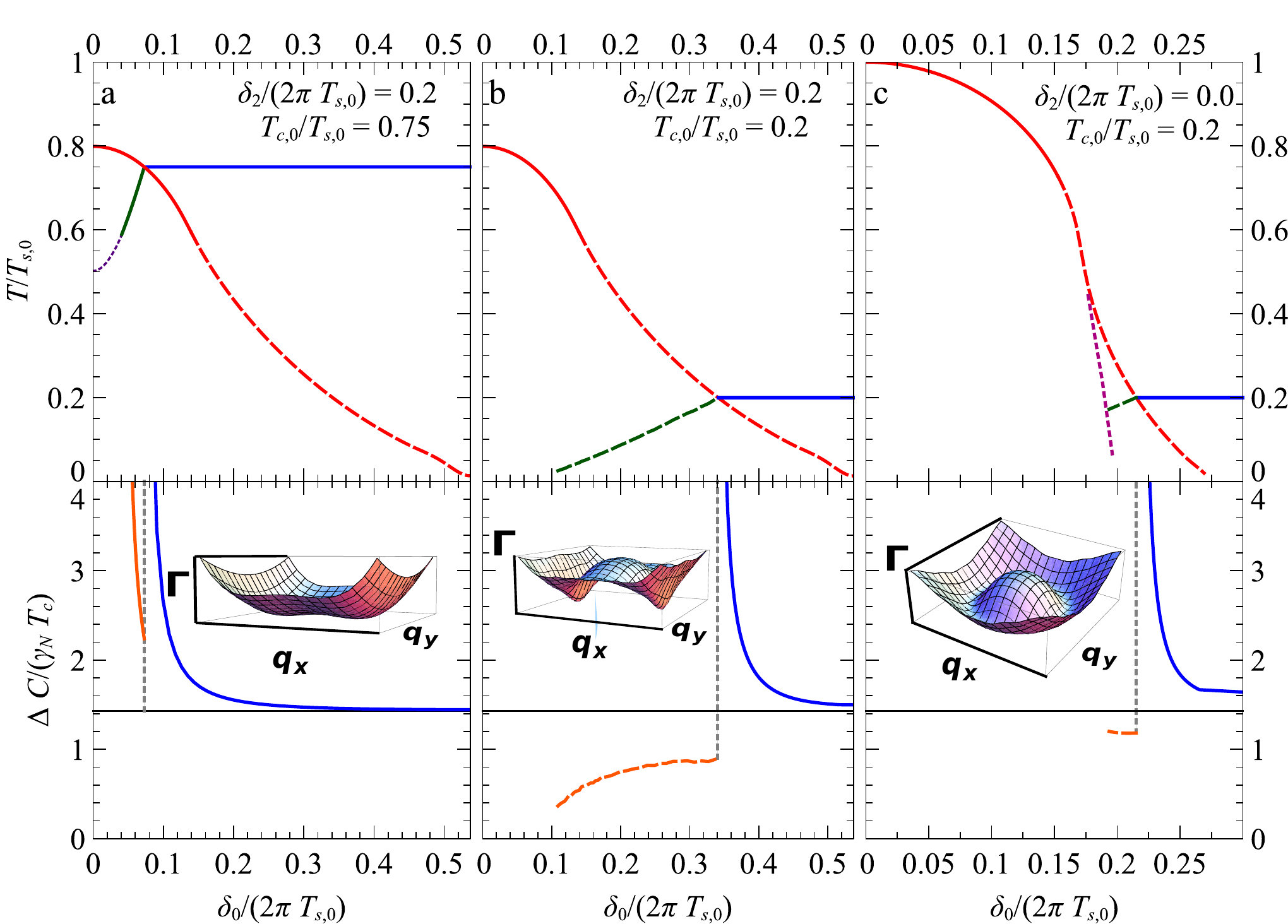}\\
  \caption{(Color online) Top: Phase diagram in $T-\delta_0$ plane
  for $\delta_{2}/(2\pi T_{s0}) = 0.2$ (panels a and b), and
  $\delta_{2}/(2\pi T_{s0}) = 0.0$ (panel c). A solid line on any
  diagram signals an SDW order parameter at a commensurate wave vector
  $\bm{Q} = \bm{\pi}$ (SDW$_0$), whereas a dashed
  line indicates incommensurate vector $\bm{Q} = \bm{\pi} + \bm{q}$
  (SDW$_{q}$) as the dominant contributor.
  Red lines indicate a second order SDW-normal phase transition.
  Horizontal blue lines correspond to the SC-normal phase transition
  temperature, which is another free parameter of the theory.
  The green lines inside the SDW phase delimit the onset of SC
  from a pre-existing SDW ordered state, ending at the tetracritical
  point at optimal doping. Purple dotted lines indicate a first-order
  phase transition between either SDW$_0$-SC phase (panel a),
  or SDW$_0$-SDW$_q$ phase (panel c).
  Bottom: Behavior of $\Delta C/T_c$ as a function of $\delta_0$
  corresponding to the situation on the top diagram in the same vertical.
  At the mean field level $\Delta C$ is discontinuous at the tetracritical
  point and jumps back to the BCS value in the overdoped region,
  which is shown by the black solid horizontal line. The mean field
  behavior in the underdoped region depends on the choice of parameters
  and may diverge if the phase transition becomes first order (as in panel a).
  Fluctuations of the SDW order parameter smear the discontinuity as shown by the blue lines.
  Insets: Behavior of $\Gamma(\bm{q})$ as defined in Eq.~(\ref{Gamma}) for the
  corresponding parameters, which determines the scaling behavior of the
  specific heat jump fluctuation correction.
\label{fig1}}
\end{figure*}

Transition temperatures from a normal phase to SDW or SC phases as
well as from SDW to the coexistence phase as functions of $\delta_0$
are depicted in upper panels of Fig.~\ref{fig1} and has been studied
in the entire range of parameters.~\cite{MV-model,FN} One spurious
property of the mean-field analysis when applied to the calculation
of the specific heat jump is apparent discontinuity of $\Delta C$
occurring when the system enters the coexistence region, see lower
panels of Fig.~\ref{fig1}. The key point to emphasize here is that
this singularity gets rounded up and transforms into a sharp peak
once we include fluctuations of the SDW order in the paramagnetic
phase. Indeed, thermodynamic fluctuations are nonzero on both sides
of the tetracritical point, and the averages
$\langle|\bm{M}^2_q|\rangle$ effectively renormalize the
superconducting part of the free energy. Similar mechanism of
enhancement of $\Delta C$ has been explored in the context of the
heavy fermion superconductors CeCoIn$_5$ and UBe$_{13}$, which
occurs due to the coupling of SC order parameter to fluctuating
magnetization of the uncompensated part of the localized $f$
moments.~\cite{Martin} However, such scenario is not directly
applicable to FeSCs since their magnetism is itinerant and spatial
fluctuations of SDW order have long correlation length. Another
important remark is that in our free energy, Eq.~\eqref{F-GL}, we
neglected gradient terms of SC order $\Delta$ since they give rise
only to subleading corrections to $\Delta C$. In other words, the
region of fluctuations is narrower for SC order than for SDW order.

\section{Renormalized Free Energy from SDW Fluctuations}

To find an effective free energy functional $\mathcal{F}(\Delta)$
near the tetracritical point, we need to integrate out magnetic
fluctuations $\bm{M}_q$ in Eq.~\eqref{F-GL}. Overdoped case $x>x_c$
differs by the absence of the finite $\langle \bm{M}_q \rangle$ from
the underdoped case $x<x_c$. However, as $\langle \bm{M}_q \rangle$
vanishes at the tetracritical point, we expect approximately the
same results in both cases. According to the general picture of
fluctuations near the second-order phase transition, we expect the
same power-exponent scaling of $\Delta C$ versus $x-x_c$ for the
underdoped and the overdoped regions of the phase diagram but with
the different pre-factors.

Expanding Eq.~\eqref{F-GL} to the leading order in $|\bm{M}_q|^2$
and performing integration at the Gaussian level we find
\begin{eqnarray}\label{F-GL-eff}
\mathcal{F}(\Delta)=-T\ln\left[\int\mathcal{D}[\bm{M}_q]\exp(-\mathcal{F}(\Delta,\bm{M}_q)/T)\right]\nonumber\\
=\mathcal{F}_{\mathrm{SC}}(\Delta)+ \delta\mathcal{F}_{\mathrm{SDW}}(\Delta).
\end{eqnarray}
The first term in the right-hand-side of Eq.~\eqref{F-GL-eff} is
simply superconducting part of the free energy which follows
directly from Eq.~\eqref{F-GL} by setting $\bm{M}_q$ and
$\delta_{\phi q}$ to zero, which thus reads
\begin{equation}
\frac{\mathcal{F}_{\mathrm{SC}}(\Delta)}{N_F}=
\frac{\Delta^2}{2}\ln\left(\frac{T}{T_{c0}}\right)- 2\pi
T\sum_{\varepsilon_n>0}\left[E_n-\varepsilon_n-\frac{\Delta^2}{2\varepsilon_n}\right].
\end{equation}
Being interested in the vicinity of the transition to the SC phase,
where SC order parameter is small, we expand renormalized free
energy $\mathcal{F}(\Delta)$ in powers of $\Delta$. The leading
order term $\mathcal{F}_{\mathrm{SC}}(\Delta)$ when expanded up to
the forth order takes the usual form for the BCS theory
\begin{equation}\label{F-SC}
\frac{\mathcal{F}_{\mathrm{SC}}(\Delta)}{N_F}=A\Delta^2+\frac{B}{2}\Delta^4
\end{equation}
with the coefficients $A=(1/2)\ln(T/T_{c0})$ and $B=(\pi
T/2)\sum_{\varepsilon_n>0}\varepsilon^{-3}_n=7\zeta(3)/16\pi^2T^2$.
From the general thermodynamic relation
$C=-T\partial^2_T\mathcal{F}$ we find from Eq.~\eqref{F-SC} that the
jump of the specific heat at the SC transition is $\Delta
C=N_FT_c[(\partial_TA)^2/B]_{T=T_c}$, which reproduces the  BCS value
$\Delta C/T_c=1.43 (\pi^2/3)N_F$.

The second term in Eq.~\eqref{F-GL-eff} is the correction to the
free energy due to SDW fluctuations
\begin{equation}\label{F-SDW}
\frac{\delta\mathcal{F}_{\mathrm{SDW}}(\Delta)}{N_F}=
\frac{3T}{2N_F}\sum_q\ln\left(\frac{K_q(T,\Delta)}{K_q(T,0)}\right),
\end{equation}
where
\begin{equation}
K_q(T,\Delta)=\ln\left(\frac{T}{T_{s0}}\right)-2\pi
T\sum_{\varepsilon_n>0}\left[ \Re \left\langle
\frac{1}{\mathcal{E}_n} \right\rangle_\phi-\frac{1}{\varepsilon_n}\right].
\end{equation}

An expansion of the correction term Eq.~\eqref{F-SDW} in powers of
$\Delta$
\begin{equation}
\label{F-SCfluct}
\frac{\delta\mathcal{F}_{\mathrm{SDW}}(\Delta)}{N_F} =\delta
A \Delta^2+ \frac{\delta B}{2} \Delta^4,
\end{equation}
leads to the renormalization of the coefficients $A$ and $B$ in
Eq.~\eqref{F-SC}, that acquire corrections
\begin{eqnarray}\label{A-B}
\delta A=\frac{3T}{2N_F}\sum_q\frac{D_1}{\Gamma},\quad \delta
B=\frac{3T}{2N_F}\sum_q\left(\frac{D_2}{\Gamma}-\frac{D^2_1}{\Gamma^2}\right).
\end{eqnarray}
The expressions that enter into Eq.~\eqref{A-B} are given by
\begin{eqnarray}
\label{Gamma}
&&\hskip-.5cm\Gamma  \equiv K_{q}(T,0)\nonumber\\
&&
\hskip-.5cm=\ln\left(\frac{T}{T_{s0}}\right)-\psi\left(\frac{1}{2}\right)+
\Re\left\langle\psi\left(\frac{1}{2}+\frac{i\delta_{\phi q}}{2\pi
T}\right)\right\rangle_{\!\!\phi},
\end{eqnarray}
\begin{equation}
\label{D1}
D_1\!=\Re\left\langle\!\frac{\psi\!
\left(\frac{1}{2}\right)\!-\psi\!\left(\frac{1}{2}+\frac{i\delta_{\phi
q}}{2\pi T}\right)}{2\delta^2_{\phi
q}}\!\right\rangle_{\!\!\phi}\!\!
-\Im\left\langle\!\frac{\psi^{[1]}\!\left(\frac{1}{2}\!+\frac{i\delta_{\phi
q}}{2\pi T}\right)}{4\pi T\delta_{\phi
q}}\!\right\rangle_{\!\!\phi}\!\!,
\end{equation}
\begin{equation}\nonumber
D_2\!=\frac{3}{4}\Re\left\langle\!\frac{\psi\!\left(\frac{1}{2}\!+\frac{i\delta_{\phi
q}}{2\pi T}\right)\!-\psi\!\left(\frac{1}{2}\right)}{\delta^4_{\phi
q}}\!\right\rangle_{\!\!\phi}\!\!\!-3\Im\left\langle\!\frac{\psi^{[1]}\!
\left(\frac{1}{2}\!+\frac{i\delta_{\phi q}}{2\pi T}\right)}{8\pi
T\delta^3_{\phi q}}\!\right\rangle_{\!\!\phi}
\end{equation}
\begin{equation}
\label{D2}
-\Re\left\langle\frac{2\psi^{[2]}\!\left(\frac{1}{2}+\frac{i\delta_{\phi
q}}{2\pi
T}\right)+\psi^{[2]}\!\left(\frac{1}{2}\right)}{32\pi^2T^2\delta^2_{\phi
q}}\right\rangle_{\!\!\phi},
\end{equation}
where $\psi$ and $\psi^{[n]}$ are the digamma and polygamma
functions respectively. Terms representing fluctuation corrections
in the free energy lead to the smearing of the discontinuity in the specific jump near
the transition $\delta(\Delta C)/\Delta C=\delta
T_c/T_c+2\partial_T\delta A/\partial_TA-\delta B/B$ with $\delta
T_c=-2T_c\delta A(T_c)$. Computing temperature derivative of the
$\delta A$ and collecting all the terms together we obtain the
following expression for the relative correction of the specific
heat jump
\begin{eqnarray}\label{delta-C}
\frac{\delta(\Delta C)}{\Delta
C}=\frac{3T_c}{2N_F}\sum_q\left[\frac{1}{\Gamma^2}\left(\frac{16\pi^2T^2_c}{7\zeta(3)}D^2_1
-4T_cD_1\partial_T\Gamma\right)\right.\nonumber\\
\left.
+\frac{1}{\Gamma}\left(2D_1+4T_c\partial_TD_1-\frac{16\pi^2T^2_c}{7\zeta(3)}D_2\right)\right],
\end{eqnarray}
which constitutes the main result of this work. To study the most
singular contribution to the specific heat jump correction in the
vicinity of the tetracritical point, we note that a second-order
transition to SDW phase is defined as the value of the detuning
parameters for which the global minimum of $\Gamma$ defined in
Eq.~(\ref{Gamma}) becomes equal to zero. Since $\Gamma$ is in the
denominator in Eq.~(\ref{delta-C}), the most singular contribution
comes from terms proportional to $1/\Gamma^2$.

\section{Scaling of $\Delta C$ near the tetracritical point}

Since the $\bm{q}$-dependence of any coefficient comes from
$\delta_{\phi q}$, some general symmetries of the function
$\Gamma(q_x, q_y; \delta_0, \delta_2)$ follow straightforwardly.
First, $\Gamma$ is symmetric in each $q$ component separately, i.e.
$\Gamma(q_x, q_y) = \Gamma(-q_x, q_y) = \Gamma(q_x, -q_y)$. Second,
a change in the sign of $\delta_2$ is equivalent to the exchange
$q_x \leftrightarrow q_y$.
We checked numerically that if there are
any local minima of $\Gamma$ at $(q_x,\pm q_y)$ for a non-zero
$q_y$, then these minima merge as $q_x$ increases until a value $q_x
= q_0$, which in this case is the point of global minimum. To model
such a behavior, it is convenient to expand $\Gamma$ in
Eq.~\eqref{Gamma} up to the fourth order in powers of $q_{x, y}$ in
the following form
\begin{equation}
\label{Gammaexp}
\begin{array}{l}
\Gamma(q_x, q_y)  \approx  A_M \\
 + \left(\begin{array}{cc} \vert q^2_x - q^2_0 \vert^m & \vert q^{2}_{y}
 \vert^n \end{array}\right) \, \left( \begin{array}{cc}
u & v \\
v & w
\end{array}\right) \, \left( \begin{array}{c}
\vert q^2_x - q^2_0 \vert^m \\
\vert q^2_y \vert^n
\end{array} \right),
\end{array}
\end{equation}
assuming that $u + w > 0$, and $u \, w - v^2 > 0$, so that both of
the eigenvalues of the matrix are positive, and $\Gamma$ is bounded
from below by $A_M$. The power exponents $m$ and $n$ are to be
chosen so that there is a quadratic dispersion around the global
minimum, unless there is a crossing from commensurate ($q_0=0$) to
incommensurate ($q_0\neq0$) SDW order, in which case quartic terms
are to be retained. In Eq.~\eqref{Gammaexp} $A_{M} = a_{M} \, (T -
T_s(x))$ near the SDW-normal phase transition. The coefficient
$a_{M} > 0$ is positive for temperatures higher than the transition
temperature and  $\Gamma > 0$. The tetracritical point is determined
for doping $x_c$ where the condition $T_s(x_c) = T_c$ is satisfied.
This leads to $A_{M} \approx -a_{M} T'_s(x_c) (x - x_c)$. The
derivative $T'_s(x_c) < 0$ is negative because doping leads to a
decrease in the SDW transition temperature. Thus, the exponent in
the scaling of the specific heat jump with $A_M$ is the same as the
exponent with $x - x_{c}$. We estimate the momentum integral in
Eq.~\eqref{delta-C} by estimating the area $S$ in the $q$-plane
where
\begin{equation}
\left(\begin{array}{cc} \vert q^2_x - q^2_0 \vert^m & \vert q^{2}_{y} \vert^n
\end{array}\right) \cdot \hat{M} \cdot \left( \begin{array}{c}
\vert q^2_x - q^2_0 \vert^m \\
\vert q^2_y \vert^n
\end{array} \right) \le A_M,
\end{equation}
with
\begin{equation}
\hat{M} \equiv \left( \begin{array}{cc}
u & v \\
v & w
\end{array}\right).
\end{equation}
We diagonalize the matrix $\hat{M}$ by an orthogonal matrix
\begin{equation}
\hat{O} = \left(\begin{array}{cc}
\cos \theta & -\sin \theta \\
\sin \theta & \cos \theta
\end{array}\right),
\end{equation}
so that $M \cdot \hat{O} = \hat{O} \cdot \hat{\Lambda}$, and
$\hat{\Lambda} = \mathrm{diag} (\lambda_1, \lambda_2)$ is a diagonal
matrix with positive eigenvalues. The angle $\theta$ is found from
the condition
\begin{equation}
\tan 2\theta = \frac{2 v}{u - w},
\end{equation}
and the eigenvalues are
\begin{equation}
\lambda_{1,2} = \frac{u + w}{2} \pm \sqrt{\left( \frac{u - w}{2} \right)^2 + v^2}.
\end{equation}
If we introduce the substitution
\begin{subequations}\label{subst1}
\begin{equation}
\vert q^2_x - q^2_0 \vert^m = \rho \, A^{1/2}_M \left[ \frac{\cos
\theta \, \cos t}{\sqrt{\lambda_1}} - \frac{\sin \theta \, \sin
t}{\sqrt{\lambda_2}} \right],
\end{equation}
\begin{equation}
\vert q^2_y \vert^n = \rho \, A^{1/2}_M \left[ \frac{\sin \theta \, \cos t}{\sqrt{\lambda_1}}
+ \frac{\cos \theta \, \sin t}{\sqrt{\lambda_2}} \right],
\end{equation}
\end{subequations}
then the region where $\Gamma \approx A_M$ is delimited by $\rho \le
1$. It is further convenient to introduce
\begin{subequations}
\begin{equation}
\frac{\cos \theta}{\sqrt{\lambda_1}} = K_x \, \cos \alpha, \quad
\frac{\sin \theta}{\sqrt{\lambda_2}} = K_x \, \sin \alpha,
\end{equation}
\begin{equation}
\frac{\cos \theta}{\sqrt{\lambda_2}} = K_y \, \cos \beta, \quad
\frac{\sin \theta}{\sqrt{\lambda_1}} = K_y \, \sin \beta.
\end{equation}
\end{subequations}
Using the definitions of $\theta$, and $\lambda_{1,2}$ from above,
we obtain
\begin{equation}
\label{coeffsK}
K^2_{x,y} = \frac{1}{u \, w - v^2} \, \left(\begin{array}{c}
w \\
u
\end{array}\right),\quad
\sin \left( \alpha - \beta \right) = \frac{v}{\sqrt{u w}}.
\end{equation}
Performing the final substitutions $t \rightarrow t - \alpha$, and
$\beta \rightarrow -\beta + \alpha$, Eq.~(\ref{subst1}), for the
first quadrant of the $q$-plane goes over to
\begin{subequations}\label{subst2}
\begin{equation}
q^{(1/2)}_x = \left[ q^2_{0} \pm \left( K_x \, \rho \, A^{1/2}_M \,
\cos t \right)^{\frac{1}{m}} \right]^{\frac{1}{2}},
\end{equation}
\begin{equation}
q_y = \left( K_y \, \rho \, A^{1/2}_M \, \sin (t - \beta)
\right)^\frac{1}{2n},
\end{equation}
\end{subequations}
where $\sin \beta$ should now stand for what was $\sin \left(\alpha - \beta \right)$, i.e.:
\begin{equation}
\label{angle} \sin \beta = \frac{v}{\sqrt{u \, w}},\qquad
-\frac{\pi}{2} < \beta < \frac{\pi}{2}.
\end{equation}
The parameter $t$ is limited by: $ \beta \le t \le \frac{\pi}{2}$,
unless $\vert q^{2}_0 \vert^ m < K_x \, A^{1/2}_M \cos \beta$. But,
since we are concerned in the regime where $A_M \rightarrow 0^{+}$,
this is only viable when $q_0 = 0$, in which case the branch
$q^{(2)}_x$ does not exit. The area $S$ of this region may be
approximated by the area of a polygon for some specific values of
the parameter $t$.

\textbf{Case-1.} When $\vert q^{2}_0 \vert^m \gg K_x \, A^{1/2}_M$
the area is approximated by the area of the triangle with vertices
at points obtained for points on the two branches $q^{(1/2)}_{x}$,
$q_y = 0$, for $t = \beta$, and the point $q^{(1/2)}_x = q_0$, and
$q_y$ for $t = \pi/2$
\begin{equation}
S\approx \frac{2}{q_0} K^{\frac{1}{m}}_x \, K^{\frac{1}{2n}}_{y}\,
\left(A_M \, \cos^2 \beta \right)^{\frac{1}{2m} + \frac{1}{4n}}.
\end{equation}

\textbf{Case-2.} If $q_0 \rightarrow 0^{+}$, then the area is
approximately that of a right-angled triangle with sides equal to
the $q_x$ and $q_y$ intercepts, obtained for $t = \beta$, and $t =
\pi/2$, respectively
\begin{equation}
S_q \approx 2 K^{\frac{1}{2 m}}_{x} \, K^{\frac{1}{2n}}_{y} \,
\left(A_M \, \cos^2 \beta \right)^{\frac{1}{4m} + \frac{1}{4n}}.
\end{equation}
The behavior of the integral is then
\begin{equation}
\iint \frac{d^2 q}{\Gamma^2(\bm{q})} \approx \frac{S}{A^{2}_M}.
\end{equation}

There is one limiting situation that arises for the choice of
parameter $\delta_2 = 0$. In that case $\Gamma$ is independent on
the azimuthal angle $\phi_0$ of the wave vector $\bm{q}$. It only
depends on the magnitude. Expanding Eq.~\eqref{Gamma} up to the
leading order around the value $q = q_0$ for which the global
minimum is obtained we get
\begin{equation}
\label{Gammaexp2} \Gamma(q) \approx A_M + u \, \vert q^2 - q^2_0
\vert^{p}
\end{equation}
where power exponent $p$ is equal to either $1$ or $2$, see table
\eqref{table1} for the summary. The integral is then easily
evaluated in the polar coordinates $\iint \frac{d^2
q}{\Gamma^2(\bm{q})} = \int_{0}^{\infty} \frac{2\pi \, q \,
dq}{\Gamma^2(q)}$. Considering two limiting cases as above, namely
$u \vert q_0 \vert^{2p} \gg A_M$, and $q_0 \rightarrow 0^{+}$, one
may show that the asymptotic behavior is as follows:
\begin{equation}
\iint \frac{d^2 q}{\Gamma^2(\bm{q})}
\approx\frac{c}{u^{1/p}}A^{1/p-2}_M,
\end{equation}
and the difference between the two cases is only in the numerical
prefactor $c=2\pi$ and $c=\frac{\pi^2(p-1)}{p^2\sin(\pi/p)}$,
respectively. A summary of the scaling exponents is given in Table~\ref{table1}.
\begin{table}[h]

\begin{tabular}{||c|c|c||}
\multicolumn{3}{c}{Anisotropic} \\
\hline
$q_0$ & power law & exp. \\
\hline
incomm. & $\vert q^2_x - q^2_0 \vert^{2m}, \; \vert q^{2}_y \vert^{2n}$ & $-2 + \frac{1}{2m} + \frac{1}{4n}$ \\

 & $2m = 2, \; 2n = 1$ & $-1$ \\
 \hline
 comm. & $\vert q^2_x \vert^{2m}, \; \vert q^2_y \vert^{2n}$ & $-2 + \frac{1}{4m} + \frac{1}{4n}$ \\
 & $2m = 1, \; 2n = 1$ & $-1$ \\
 & $2m = 2, \; 2n = 1$ or $2m = 1, \; 2n = 2$ & $-5/4$ \\
 \hline\hline
 \multicolumn{3}{c}{} \\
 \multicolumn{3}{c}{Isotropic} \\
 \hline
$q_0$ & power law & exp. \\
\hline
 incomm. & $\vert q^2 - q^2_0 \vert^p$ & $-2 + \frac{1}{p}$ \\
  & $p = 2$ & $-3/2$ \\
\hline
comm. & $ \vert q^2 \vert^p$ & $-2 + \frac{1}{p}$ \\
 & $ p = 1$ & $-1$ \\
\hline\hline
\end{tabular}
\caption{\label{table1} Typical exponents for the scaling behavior
of the most singular fluctuation correction for the specific heat
jump. ``incomm.'' denotes that the global minimum is realized for
$q_0 \neq 0$, while ``comm.'' that it is for $q_0 = 0$.}
\end{table}

\begin{figure}
  \includegraphics[width=8cm]{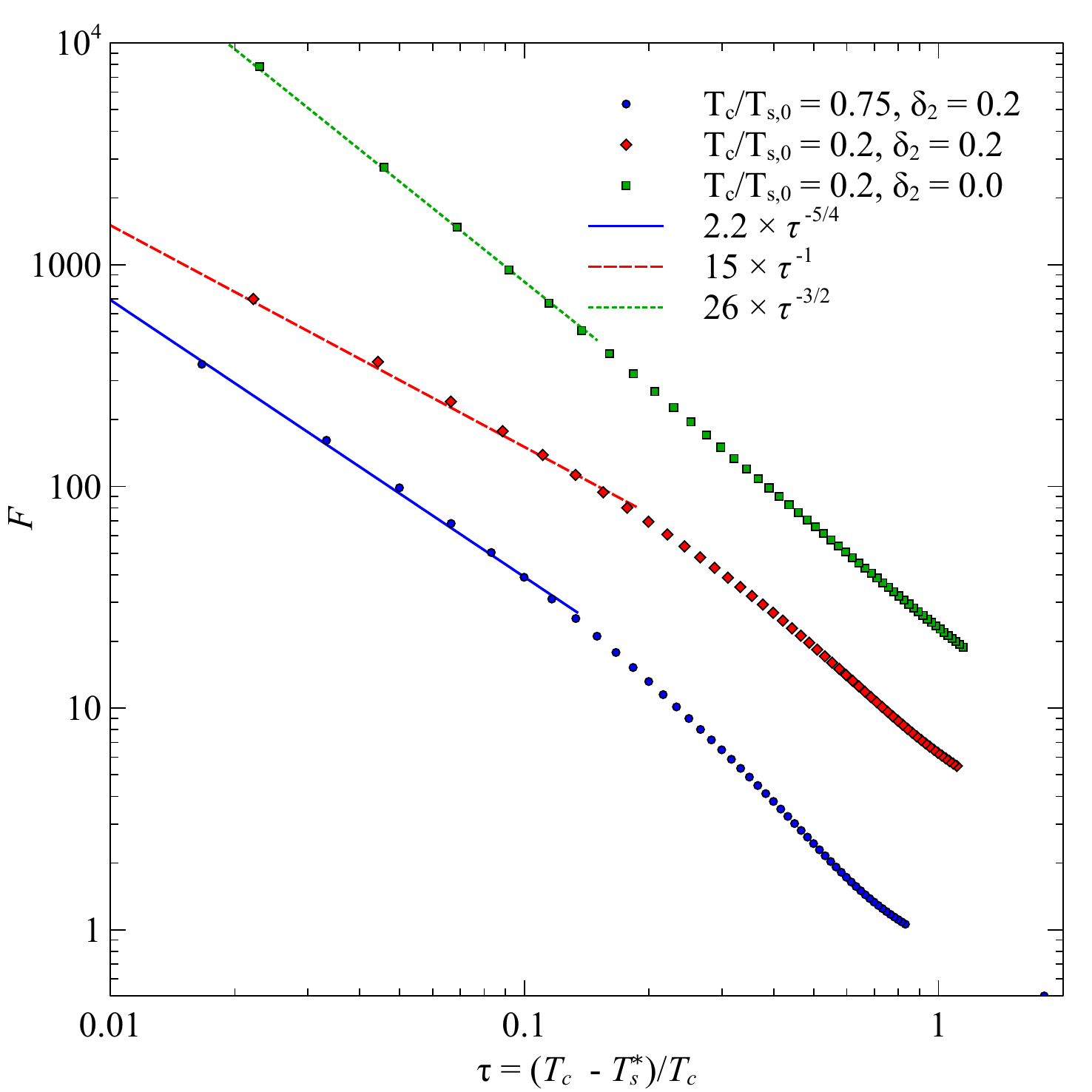}\\
  \caption{(Color online) A log-log plot of the most singular specific
  heat jump fluctuation correction. The exponent varies between -1 and -2.
  The deviation from power law dependence for large values of $\tau$
  is due to inessential band structure effects for the topic at hand.}
\label{fig2}
\end{figure}

Having above analytical arguments, we evaluate correction to the
specific heat jump numerically and present our results in
Fig.~\ref{fig2} on a log-log plot for the same choice of parameters
as in Fig.~\ref{fig1}.
A useful dimensionless parameter that characterizes the deviation
from the tetracritical point, and one that is customarily chosen is
\begin{equation}
\label{tauparameter} \tau \equiv \frac{T_{c} -
T^{\ast}_{s}(\delta_0)}{T_c}.
\end{equation}
Here $T^{\ast}_{s}(\delta_0) = T_{s}(\delta_{0c}) +
T'_{s}(\delta_{0c})(\delta_0 - \delta_{0c})$ is the linearized SDW
transition temperature dependent on the isotropic detuning parameter
$\delta_0$ near the tetracritical point, defined as
$T_{s}(\delta_{0c}) = T_{c}$. We use the linearized temperature
dependence in order to cancel any additional power law scaling
coming from the nonlinear dependence. Assuming that $\delta_0$ is a
linear function of $x$, this enables us to study scaling in terms of
experimentally measurable $x - x_c$. In this way, we obtain the
following numerical law
\begin{equation}
\label{Cnumeric} \frac{\delta (\Delta C)_{\mathrm{sing.}}}{\Delta C}
= \kappa \, F\left(\tau, \delta_2, \frac{T_{c0}}{T_{s0}}\right),
\end{equation}
where $\kappa$ is a dimensionless combination of several constants
characterizing the system
\begin{equation}
\label{smallparam} \kappa = \frac{6}{\pi^2} \, \frac{T_{c0}}{N_F \,
v^2_F} = \frac{3}{2\pi} \frac{T_{c0}}{T_{F}}.
\end{equation}
In the last step, we used the fact that for a parabolic dispersion
in 2-D, $N_F \, v^{2}_F = 2b T_{F}/\pi$, where $b$ is the number of
FS pockets (in our case $b = 2$.) This prefactor plays the role of a
small parameter in our approximation scheme.  When $\kappa F \sim
1$, the contribution to the specific heat jump from fluctuations
becomes comparable to the mean field contribution, indicating that
the logarithmic derivative approximation to derive the correction
Eq.~(\ref{delta-C}) becomes inapplicable. The fact that the ratio
$T_{c0}/T_F$ takes a numerical value of the order of $10^{-2}$ in
the iron-pnictide compounds, limits the validity of the correction
to lower values of $\tau$ in the region of $0.05 - 0.5$, while
effects of details in the band structure certainly become prominent
when $\tau \sim 1$.

If one performs a similar analysis for the sub-dominant term in
Eq.~(\ref{delta-C}), one would naturally obtain a correction that
scales logarithmically $\propto\ln(x-x_c)$ for the most typical
$q$-dependence of $\Gamma$. Since our analysis dealt with an
effective action for the two order parameters from the very
beginning, we interpret this correction as arising due to the SDW
fluctuation correction of the two-point correlator of the SC order
parameter. Drawing skeleton diagrams with explicit appearance of
fermion lines and SC-fermion interaction vertices, and ``dressing
them'' with SDW fluctuations, one sees that, aside from the
self-energy (mass) renormalization of the fermions, there are also
vertex-correction contributions. Furthermore, there are
contributions from the four-point SC correlator that involve even
more complicated constructs in terms of language of fermions and our
analysis captures all these effects.

\section{Comparison to Experiment}
\begin{figure}
\includegraphics[width=8.5cm]{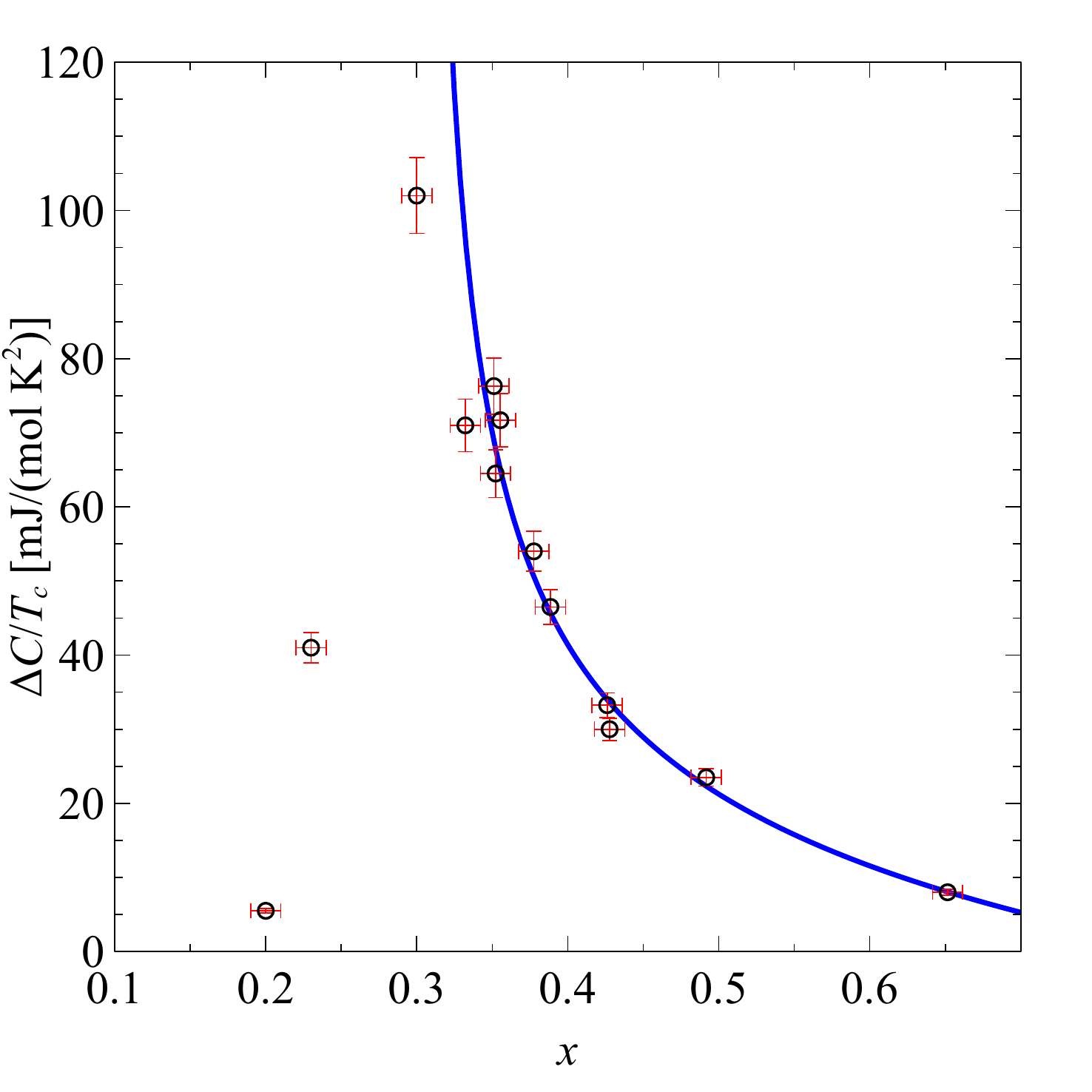}
  \caption{\label{fig3} The size of the jump in the specific heat
as a function of doping. Points with error bars represent
experimental data of Ref.~\onlinecite{QCP-C}. The solid line shows a
combined fit that includes a contribution from fluctuations of SDW
order. The left three points correspond to underdoped samples and
are not taken into account for fitting curves.}
\end{figure}

We perform  a weighted  least-squares fit, including the combined
errors in $\Delta C/T_c$, as well as $x$,  of the model
\begin{equation}
\label{model-fit}
\frac{\Delta C}{T_c} = \alpha +\beta \, \ln(x-x_c)+ \gamma \, (x-x_c)^{-1},
\end{equation}
to data from recent experiment of Walmsley \textit{et
al.}~\cite{QCP-C}, using ten points
for overdoped samples,  as shown in Fig.~\ref{fig3}.
Eq.~(\ref{model-fit}) incorporates a constant
(BCS), and a logarithmically dependent (``QCP'') term, as well as
the most generic power-law dependence $(x - x_c)^{-1}$ found in the
cases enumerated in  Table~\ref{table1}. The critical doping $x_c =
0.3$ corresponds to the optimally doped sample, and is  held fixed.
The fit gave the following values
\begin{equation}
\label{fit2} \alpha = -14.4, \quad \beta = -16.9, \quad \gamma =
1.7,
\end{equation}
with a reduced chi-squred $\chi^2_\nu = 0.91$ for $\nu = 10 - 3 = 7$
degrees of freedom. From these numerical values, one can conclude
that the power-law contribution is larger than the logarithmic term
for doping $x_c < x < x_c + 0.03$.

\section{Conclusions}

We have studied doping dependence of the specific heat jump in FeSCs
based on a minimal two-band model of electron band structure. We
have found that beyond the mean field level the discontinuity of
$\Delta C/T_c$ at the tetra-critical point (the end point of the
coexistence phase) transforms into the sharp maximum. As a result,
$\Delta C/T_c$ drops for deviations from $T_c(x_c) = T_s(x_c)$ both
into the coexistence phase and away from the SDW region. Still, the
decrease of $\Delta C/T_c$ should be more rapid within the
SDW-ordered phase. In the vicinity of the optimal doping $x_c$ the
scaling of $\Delta C/T_c$ versus $x-x_c$ is governed by the two main
effects. The first is logarithmic quasiparticle mass renormalization
which stems from the quantum critical fluctuations beneath
superconducting dome.\cite{AL-Lambda,Debanjan-Lambda,Takuya-Lambda}
The second is the effect of thermal fluctuations of the SDW order.
Our numerical fitting procedure to the data of
Ref.~\onlinecite{QCP-C} suggests the significant importance of the
thermal SDW fluctuations on the magnitude of the specific heat jump
at the transition to the SC phase.

\subsection*{Acknowledgments}

We thank A.~Chubukov, Y.~Matsuda and T.~Shibauchi for discussions. We are
especially grateful to A.~Carrington for numerous discussions, for
reading the manuscript, and for providing us with experimental data,
which enabled us to perform the comparison presented in Fig.~3. A.L.
acknowledges support from NSF under Grant No. PHYS-1066293 and the
hospitality of the Aspen Center for Physics where part of this work
was performed. M.K. acknowledges support from University of Iowa.
D.K and M.V. were supported by NSF Grant No. DMR 0955500.


\begin{thebibliography}{99}
\bibitem{FN}
We note that the SDW-normal phase transition curve in
Figs.~\ref{fig1}a, b differs from the one in
Ref.~\onlinecite{MV-model} [see their Fig.~5], while
Fig.~\ref{fig1}c is identical. The reason for this is that in the
presence of anisotropic detuning, there is partial compensation of
the detrimental effects of anisotropy and incommensurability for the
survival of the SDW phase, extending the SDW$_{q}$ phase above the
critical value $\delta_{0c} = e^{-\gamma_E}/2 = 0.2807$ valid in the
case when either $\delta_2 = 0$, or $q = 0$.
\end{thebibliography}
\end{document}